\documentclass[a4paper]{article}
\usepackage[cp1250]{inputenc}
\usepackage[T1]{fontenc}
\newcommand{\bd}{\begin{displaymath}}
\newcommand{\ed}{\end{displaymath}}
\newcommand{\ee}{\end{equation}}
\newcommand{\be}{\begin{equation}}
\newcommand{\la}{\lambda_1}
\newcommand{\lb}{\lambda_2}
\newcommand{\lc}{\lambda_3}
\newcommand{\ld}{\lambda_4}
\newcommand{\lp}{\lambda_5}

\newcommand{\lczp}{\lambda_{345}}
\newcommand{\ba}{\begin{array}}
\newcommand{\ea}{\end{array}}
\newcommand{\g}{\,\mbox{GeV}}

\usepackage[pdftex]{graphicx}
\usepackage{amsmath}
\usepackage{amssymb}
\usepackage{amsthm}
\usepackage[lofdepth,lotdepth]{subfig}
\title{Tree-level unitarity constraints for the SM-like 2HDM}
\author{Bogumi\l a Gorczyca and Maria Krawczyk\\
{\it{Faculty of Physics, University of Warsaw}}}
\begin{document}
\maketitle
\setlength\arraycolsep{2pt}
\abstract{We consider a $\mathbb{Z}_2$-symmetric Two Higgs Doublet Model (2HDM), where the vacuum expectation values of both doublets are non-vanishing. Unitarity constraints for the SM-like 2HDM are analyzed for the mass of the lightest Higgs boson in the range $[115,\,127]\g$. New stringent bounds on $\tan\beta$, $0.17\leqslant\tan\beta\leqslant6.10$, are derived. A discussion of the source of these constraints is provided.}


\section{Introduction}
Unitarity constraints for the Two Higgs Doublet Model (2HDM) have been extensively studied before \cite{huffel, weldon, casa, casal, malampi, kanemura, akeroyd, ivanov, gi, nowa}, however the SM-like limit of the model has been considered only in the case with violation of $\mathbb{Z}_2$ symmetry~\cite{tsu}. Bounds on the scalar particles' masses following from the basic assumption of tree-level unitarity in different versions of 2HDM, including the Inert Model, have been considered recently by one of us~\cite{ja}.
Here we present an extension of this analysis for the SM-like 2HDM. We assume that the mass of the lightest Higgs boson is in the range $[115,\,127]\g$ in agreement with the newest results from the LHC for the SM Higgs boson presented in December 2011~\cite{lhc}. 
\section{The general and the  SM-like Mixed Models}

We consider a 2HDM with a general $\mathbb{Z}_2$-symmetric potential for the $SU(2)$ doublets $\phi_S$, $\phi_D$ \cite{kr}:\renewcommand{\arraystretch}{1.5}
\be\label{pot}
\renewcommand{\arraystretch}{1.5}
\begin{array}{rcl}
V&=&-\frac{1}{2}\Big[m_{11}^{2}(\phi_{S}^{\dagger}\phi_{S})+m_{22}^{2}(\phi_{D}^{\dagger}\phi_{D})\Big]+\frac{1}{2}\Big[\lambda_{1}(\phi_{S}^{\dagger}\phi_{S})^{2}+\lambda_{2}(\phi_{D}^{\dagger}\phi_{D})^{2}\Big]+\\
&+&\lambda_{3}(\phi_{S}^{\dagger}\phi_{S})(\phi_{D}^{\dagger}\phi_{D})+\lambda_{4}(\phi_{S}^{\dagger}\phi_{D})(\phi_{D}^{\dagger}\phi_{S})+\frac{1}{2}\lambda_{5}\Big[(\phi_{S}^{\dagger}\phi_{D})^{2}+(\phi_{D}^{\dagger}\phi_{S})^{2}\Big].\\
\end{array}
\ee
The parameters $m_{11}^{2}$, $m_{22}^{2}$ and $\la\ldots\ld$ are real numbers and without loss of generality \cite{kr, cp, sym} we can assume that $\lp$ is real. For a stable vacuum state to exist it is necessary that \cite{ma}:
\be\label{pos}
\textrm{a)}\; \la>0,\quad\lb>0,\quad \textrm{b)}\;\lc+\sqrt{\la\lb}>0,\quad \textrm{c)}\;\lczp+\sqrt{\la\lb}>0,
\ee
where $\lczp=\lc+\ld+\lp$.

We consider a model (called a Mixed Model) in which a Mixed vacuum is realized, i.e. in the state of the lowest energy (the global minimum of the potential) both of the doublets develop non-zero vacuum expectation values: $\langle\phi_S\rangle=v_S/\sqrt{2}$, $\langle\phi_D\rangle=v_D/\sqrt{2}$, $v_D / v_S=\tan\beta$, $\beta\in(0,\,\pi/2)$. The Mixed vacuum exists when the following conditions are satisfied~\cite{kr, ja}
\be
\label{mix}
v_{S}^{2}=\frac{m_{11}^{2}\lb-\lczp m_{22}^{2}}{\la\lb-\lczp^{2}}>0,\quad    v_{D}^{2}=\frac{m_{22}^{2}\la-\lczp m_{11}^{2}}{\la\lb-\lczp^{2}}>0,\quad
\ld+\lp<0,\quad \lp<0,\quad \la\lb-\lczp^{2}>0.
\ee\renewcommand{\arraystretch}{1}
The matrix of second derivatives of the potential is non-diagonal, so the mass eigenstates are mixtures of the fields appearing in the potential.  There is one mixing angle $\beta$ in the charged  and  the CP-odd sectors and  one  mixing angle $\alpha$ in the CP-even sector, $\alpha\in(-\pi/2,\,\pi/2)$.
There arise 3 Goldstone bosons and 5 physical Higgs particles: $H^{\pm}$, $A$, $H$, $h$, with $M_H>M_h$. Different models of Yukawa interactions can be chosen
, however  we do  not fix a particular one here, because Yukawa interactions do not effect the following analysis.

When an additional condition:
\be\label{sm-like}
\sin(\beta-\alpha)=1
\ee
is imposed, $h$  couples to gauge bosons at the tree-level  like the SM Higgs particle\footnote{And so it does to fermions, if e.g. Model II of Yukawa interactions is chosen.} \cite{sm-like}. Then, it is justified to apply  the following experimental bounds  on its mass,
\be
\label{Mhexp}
M_h\in[115,\,127]\g,
\ee
found for the SM Higgs boson \cite{lhc}. A Mixed Model with additional assumptions (\ref{sm-like}) and (\ref{Mhexp}) we call  a SM-like Mixed Model.
In the following we will focus on this model,
however, for   comparison  results for the general Mixed Model are  presented as well.


\section{Tree-level unitarity approach}
To derive constraints from unitarity we follow the standard high-energy approach, where longitudinally polarized states of vector bosons are replaced by the corresponding would-be Goldstone bosons \cite{lqt} and only the quartic interactions are included in the $2\to2$ scatterings. In this analysis we consider the scattering matrix for the original fields of the Lagrangian, i.e. not mass eigenstates, as this makes analysis simpler~\cite{kanemura}.

Inequalities resulting from the unitarity condition on the s-wave
\be\label{unit}
|\Re(a^{(0)})|<\frac{1}{2}
\ee
are considered. They are inferred from the full tree-level high-energy scattering matrix of the scalar sector with 25 different channels \cite{ja}. The inequalities are solved numerically (as in \cite{akeroyd}) taking into account explicitly  the positivity constraints (\ref{pos}) and conditions necessary for the existence of Mixed vacuum (\ref{mix})\footnote{It is not obvious if they were taken into account in some earlier analyses.}, as well as 
two additional constraints (\ref{sm-like}) and (\ref{Mhexp}) for the SM-like Mixed Model.
For simplicity the set of the relevant conditions, namely: tree-level unitarity constraint (\ref{unit}) together with constraints (\ref{pos}) and (\ref{mix})   (for the SM-like Model the set contains in addition two constraints (\ref{sm-like}) and (\ref{Mhexp})) will be called the unitarity constraints.

A scan, subject to the unitarity constraints, over different values of $\tan\beta$ and Higgs boson masses is performed in the following ranges:
\be\label{range}
\tan\beta\in[0,\,60]\footnote{Although, there exists a lower bound on $\tan\beta$ which comes from the assumption of perturbativity of the $\overline{t}bH^{\pm}$ coupling and is valid in the models I-IV of Yukawa interactions~\cite{tanb}, we do not impose it not to obscure the effects of the considered constraints.},\,
\,  M_A,\,M_H,\,M_{H^{\pm}}\in[0,\,720]\g,\, M_h\in[0,\,M_H].
\ee


\section{Results}
We perform a scan for the general and the SM-like Mixed Models and as a result
bounds on scalars' masses
as well as on $\tan \beta$ are obtained.

\subsection{General Mixed Model}

Unitarity constraints lead to the following upper bounds on the Higgs bosons' masses in the general Mixed Model:
\be\label{eq:mixed-wynik}
\begin{array}{r@{\;\leqslant\;}l}
M_{H^{\pm}}&690\,\textrm{GeV},\\
M_{A}&711\,\textrm{GeV},\\
M_{H}&688\,\textrm{GeV},\\
M_{h}&499\,\textrm{GeV}.\\
\end{array}
\ee
Note that $h$ is remarkably lighter than the other Higgs bosons and that $\tan\beta$ is not bounded by the unitarity constraints.
These results agree at the  level of 1-3\% with the most precise analytical results obtained in Ref.\,\cite{nowa}.

\subsection{SM-like Mixed Model}

\begin{figure}[h]
\centering
\includegraphics[width=0.33\textwidth]{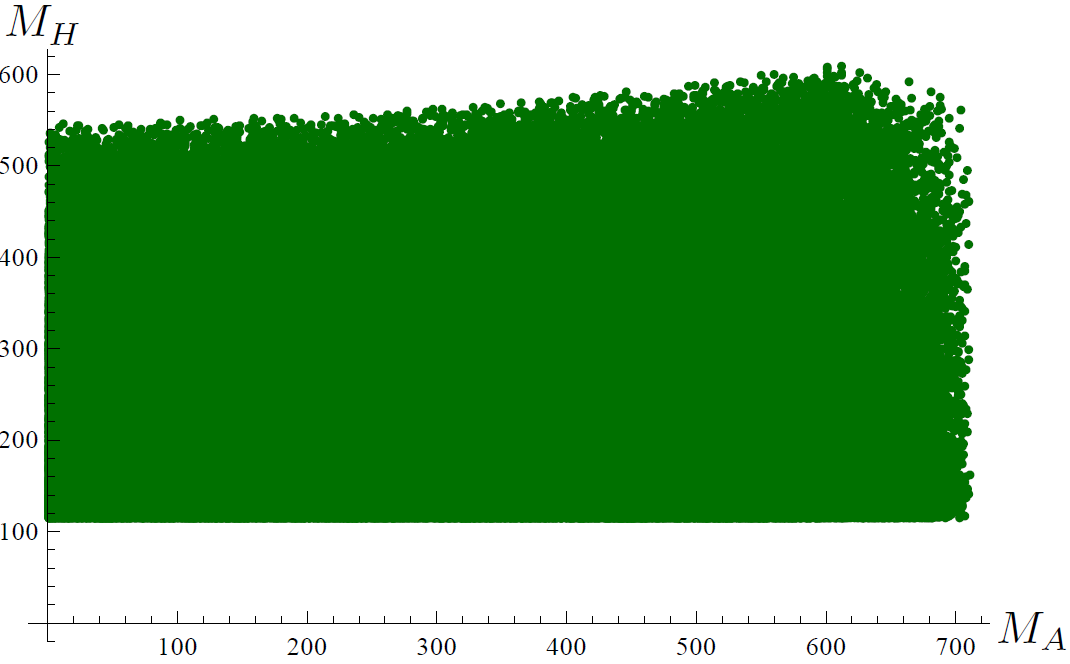}
\includegraphics[width=.33\textwidth]{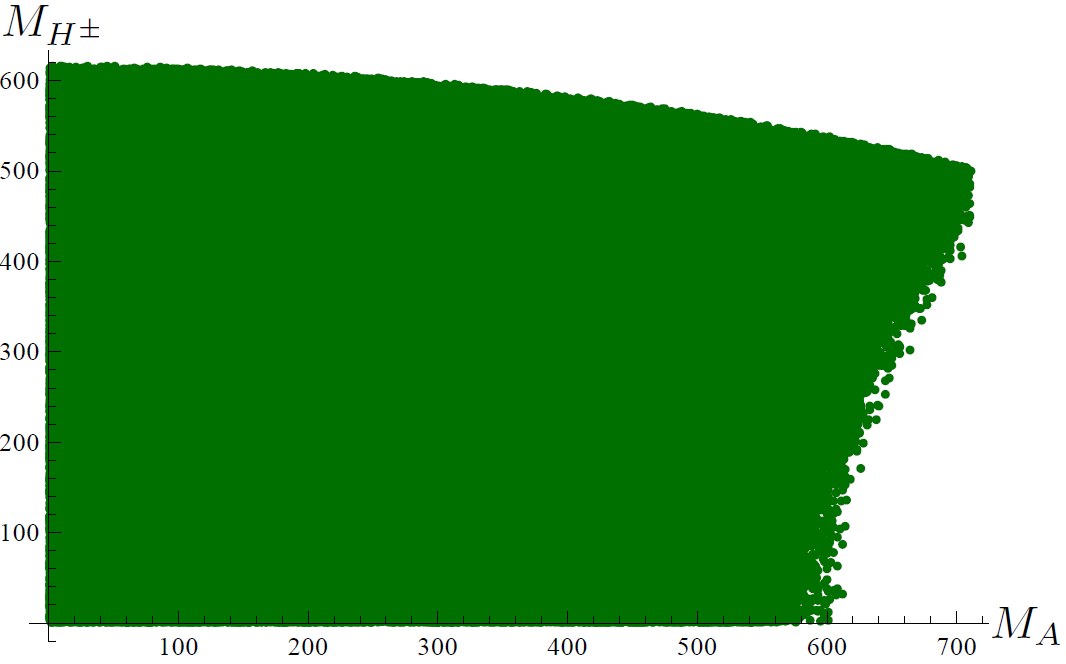}
\includegraphics[width=.33\textwidth]{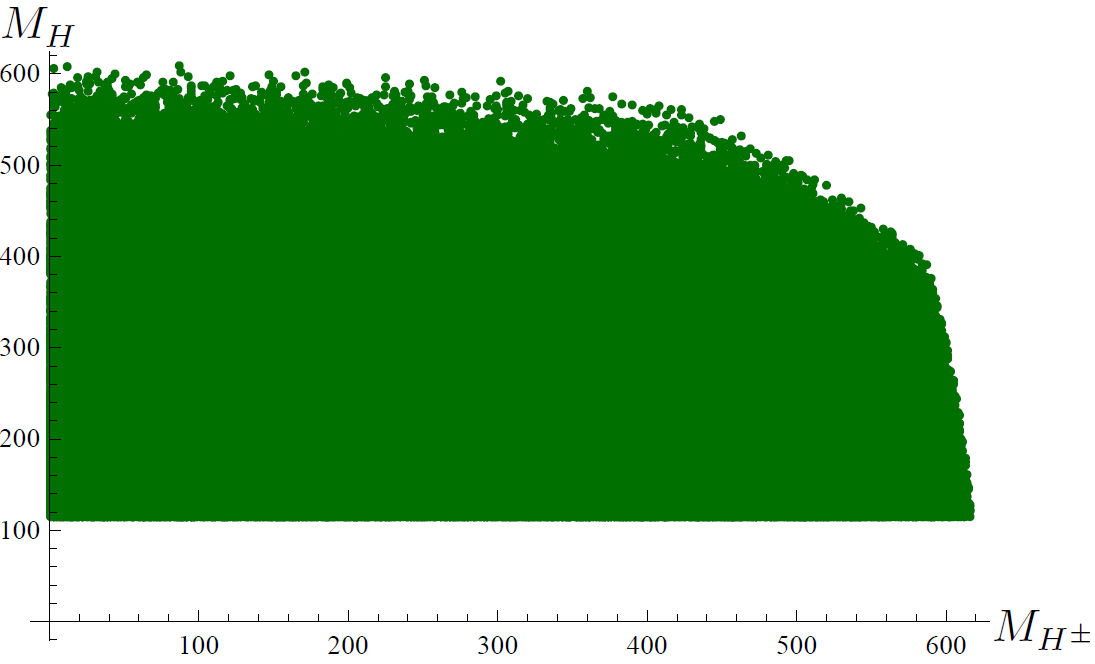}
\caption{Regions of masses allowed in the SM-like Mixed Model by the unitarity constraints. $M_h$ does not display significant correlations with masses of other Higgses, thus corresponding plots are not shown. 
\label{masy}}
\end{figure}

\begin{figure}[ht]
\centering
\includegraphics[width=.5\textwidth]{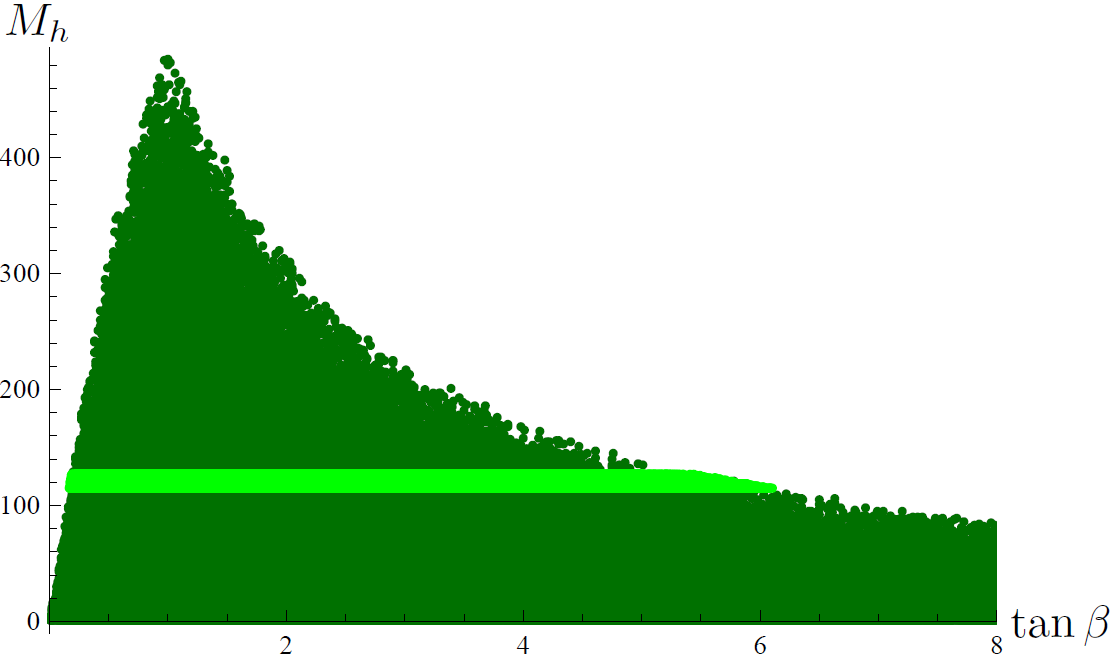}
\includegraphics[width=0.5\textwidth]{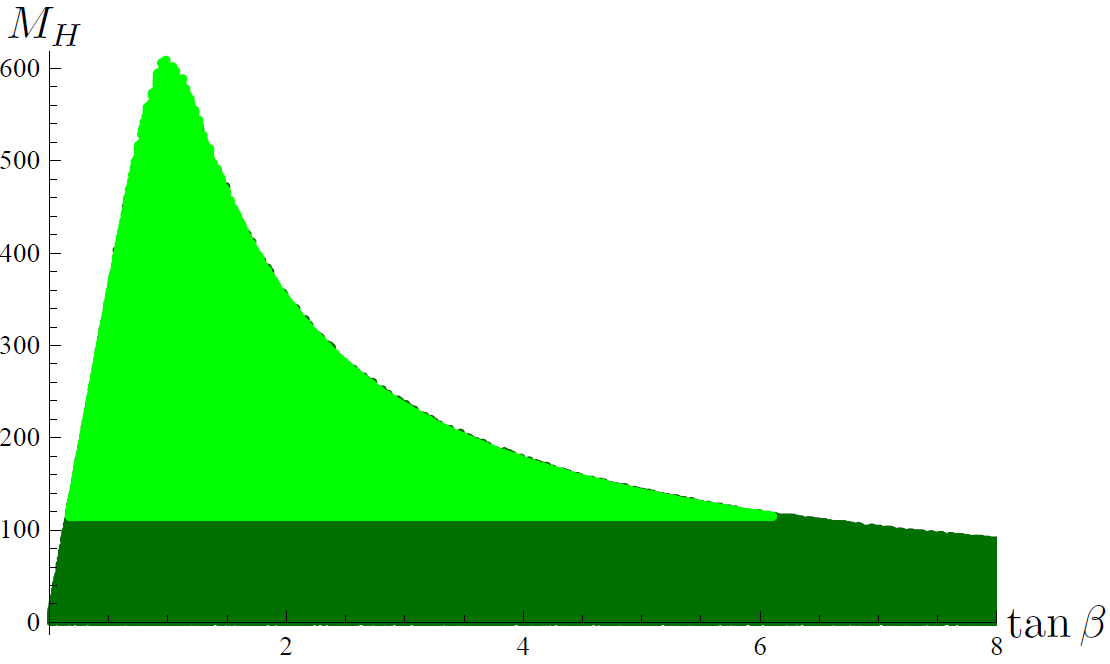}
\caption{Regions of masses:  $M_h$ (left) and $ M_H$ (right) versus $\tan\beta$  allowed in the SM-like Mixed Model
by the unitarity constraints 
(pale points). For comparison results of a scan without experimental bounds imposed on $M_h$ are presented (dark  points). The remaining regions of allowed masses do not exhibit dependence on $\tan\beta$. \label{masytb}}
\end{figure}

For the SM-like Mixed Model we perform an analysis analogous to the one for the general case discussed above.    The following upper bounds are found:
\be\label{eq:bound}
\begin{array}{r@{\;\leqslant\;}l}
M_{H^{\pm}}&616\,\textrm{GeV},\\
M_{A}&711\,\textrm{GeV},\\
M_{H}&609\,\textrm{GeV},\\
\end{array}
\ee and 
\be\label{tb}
0.17\leqslant\tan\beta\leqslant6.10.
\ee
We see that the bounds for $H^{\pm}$ and $H$ are lowered by 70-80$\g$ in comparison to the general case (compare with Eq.\,(\ref{eq:mixed-wynik})). What definitely draws attention is the stringent bound on $\tan\beta$ which arises in this case. In the next section we discuss  main sources of this strong bound. The regions  of masses allowed by the unitarity constraints and their correlations with $\tan\beta$ are presented in Fig.\,\ref{masy} and Fig.\,\ref{masytb} (pale area), respectively. The upper bounds, Eqs.~(\ref{eq:mixed-wynik}) and (\ref{eq:bound}), should be treated as maximal values  reached during the scan. It should be underlined that these values not always can be approached simultaneously, as it can be inferred from the  Figs. \ref{masy} and \ref{masytb}, where allowed regions in respective parameter spaces are presented.

The bounds presented above suffer from some uncertainties inherent to the numerical method in use. The uncertainties can be evaluated using Figs. 1 and 2. If the boundaries of the regions of allowed masses are smooth and well filled with points the uncertainties are small (should not exceed 1\%), while  if the boundaries consist of separated points, the uncertainties are obviously larger\footnote{They can be estimated as being of order of the distances between nearest points divided by the upper bound on respective mass.}.

\section{Discussion of the bound on $\tan\beta$ in the SM-like Mixed Model}

\subsection{The role of the experimental bound on $M_h$}
The correlations between the masses and $\tan\beta$ allowed by the unitarity constraints in the SM-like Mixed Model are  presented in Fig.\,\ref{masytb} (pale area). For comparison analogous results obtained without condition (\ref{Mhexp}) imposed are presented (Fig.\,\ref{masytb}, dark area)\footnote{The large  $\tan\beta$ regions have not been included in the plot, because  the maximal allowed Higgs boson masses monotonically fall down with raising $\tan\beta$ to approach the values $M_h\approx M_H\approx12\g$.}.

The mass of $h$ can be expressed in the Mixed Models  as~\cite{kr}:
$$
M_h^2=\frac{v^{2}}{2}\frac{1}{1+\tan^{2}\beta}\Big(\la+\lb\tan^{2}\beta-\sqrt{(\la-\lb\tan^{2}\beta)^{2}+4\lambda_{345}^{2}\tan^{2}\beta}\Big).
$$
Hence for fixed $\lambda$'s we have
\be\label{limtb1}
M_{h}^{2}\to\frac{v^{2}}{2}\Big(\lb-\sqrt{\lb^{2}}\Big)=0\;\mbox{for}\; \tan\beta\to\infty \quad
\textrm{and}\quad
M_{h}^{2}\to\frac{v^{2}}{2}\Big(\la-\sqrt{\la^{2}}\Big)=0\;\mbox{for}\;\tan\beta\to0.
\ee
Therefore $\tan\beta$ cannot be neither too small nor too large if  the lower limit on $M_h$ is applied. In general, an upper bound on $\tan\beta$ will arise whenever a lower bound on $M_h$ is imposed.

 In Fig.\,\ref{masytb} (left panel, pale area)
the allowed region  for $M_h$ and $\tan \beta$ is shown.  It is clear that the upper and lower limits on  $\tan \beta$ are correlated with lower limit  $M_h=114\g$, the same holds for $M_H$ (Fig.\,\ref{masytb}, right panel, pale area). 

These results are in agreement with the reasoning (\ref{limtb1}) and confirm that the bounds on $\tan\beta$,  Eq.\,(\ref{tb}), are due to the experimental lower bound on $M_h$ mass (\ref{Mhexp}).


\subsection{The role of the condition $\sin(\beta-\alpha)=1$}

If we waive the constraint (\ref{sm-like}) $\sin(\beta-\alpha)=1$ but keep the experimental limits (\ref{Mhexp}) on $M_h$, the  bounds on $\tan\beta$ are found  to be almost the same as in Eq.\,(\ref{tb}), namely $0.17\leqslant\tan\beta\leqslant6.11$. Thus, the bounds on $\tan\beta$ are arise mainly due to the constrain on $M_h$ and it is not necessary that $h$ couples to gauge bosons and fermions like the SM Higgs to obtain them.


\section{Discussion and summary}

We considered the tree-level unitarity for  the SM-like Mixed Model, and found very stringent bounds on $\tan\beta$: $0.17\leqslant\tan\beta\leqslant6.10$. These bounds   have been shown to be induced by the lower experimental limit (115 GeV) imposed on $M_h$. It is worth noticing that the limits on $\tan\beta$ are obtained without specifying the type of the Yukawa interactions and without applying constraints \textit{directly} on couplings with gauge bosons (i.e. on $\sin(\beta-\alpha)$).

\section*{Acknowledgments}
We are grateful to  D.~Sokołowska, P.~Chankowski, G.~Gil, I.~F.~Ginzburg, K.~A.~Kanishev and H.~Haber for fruitful discussions. 
Work was partly supported by Polish Ministry of Science and Higher Education Grant N N202 230337.


\begin{thebibliography}{99}

\bibitem{huffel} H.~H\"uffel, G.~P\'ocsik, \emph{Unitarity Bounds on Higgs Boson Masses in the Weinberg-Salam Model with Two Higgs Doublets}, Z.~Phys.~C - Particles and Fields 8 (1981) 13-15.
\bibitem{weldon} H.~A.~Weldon, \emph{Effects of multipleHiggs bosons on tree unitarity}, Phys.~Rev.~D 30 (1984) 1547-1558.
\bibitem{casa} R.~Casalbuoni, D.~Dominici, R.~Gatto, C.~Giunti, \emph{Strong interacting two-doublet and doublet-singlet Higgs models}, Phys.~Lett. B 178 (1986) 235-240.
\bibitem{casal} R.~Casalbuoni, D.~Dominici, F.~Fergulio, R.~Gatto, \emph{Tree-level unitarity violation for large scalar mass in multi-Higgs extensions of the Standard Model}, Nucl.~Phys.~B 299 (1988) 117-150.
\bibitem{malampi} J.~Maalampi, J.~Sirkka, I.~Vilja, \emph{Tree level unitarity and triviality bounds for two-Higgs models}, Phys.~Lett.~B 265 (1991) 371-376.
\bibitem{kanemura} S.~Kanemura, T.~Kubota, E.~Takasugi, \emph{Lee-Quigg-Thacker bounds for Higgs boson masses in a two-doublet model}, Phys.~Lett.~B 313 (1993) 155-160.
\bibitem{akeroyd} A.~G~Akeroyd, A.~Arhrib, E.~Naimi, \emph{Note on tree-level unitarity in the General Two Higgs Doublet Model}, Phys.~Lett.~B 490 (2000) 119-124;\\
A.~Arhrib, \emph{Unitarity constraints on scalar parameters of the Standard and Two Higgs Doublets Model}, Proceedings of ``Noncommutative Geometry, Superstrings and Particle Physics'' Workshop, Univ-Rabat, Morocco, June 16-17 2000, arXiv:hep-ph/0012353v1.
\bibitem{ivanov} I.~F.~Ginzburg, I.~P.~Ivanov, \emph{Tree level unitarity constraints in the 2HDM with CP violation}, arXiv:hep-ph/0312374v1.
\bibitem{gi} I.~F.~Ginzburg, I.~P.~Ivanov, \emph{Tree-level unitarity constraints in the most general 2HDM}, Phys.~Rev.~D 72 (2005) 115010.
\bibitem{nowa} J.~Ho\v{r}ej\v{s}\'i, M.~Kladiva, \emph{Tree-unitarity bounds for THDM Higgs masses revisited}, Eur.~Phys.~J. C 46 (2006) 81.
\bibitem{tsu} S.~Kanemura, Y.~Okada, H.~Taniguchi, K.~Tsumura, \emph{Indirect bounds on heavy scalar masses of the two-Higgs-doublet model in light of recent Higgs boson searches}, Phys.~Lett.~B 704 (2011) 303.
\bibitem{ja} B.~Gorczyca, \emph{Unitarity constraints for the Inert Doublet Model}, Master Thesis at the University of Warsaw (2011), in Polish.


\bibitem{lhc} The report of ATLAS experiment from 13.12.2011: http://www.atlas.ch/news/2011/status-report-dec-2011.html;\\
The report of CMS experiment from 13.12.2011:http://cms.web.cern.ch/news/cms-search-standard-model-higgs-boson-lhc-data-2010-and-2011.

\bibitem{kr} I.~F.~Ginzburg, K.~A.~Kanishev, M.~Krawczyk, D.~Soko\l owska, \emph{Evolution of Universe to the present inert phase}, Phys.~Rev.~D 82 (2010) 123533.

\bibitem{cp} G.~C.~Branco, L.~Lavoura, J.~P.~Silva, \emph{CP Violation}, 1999, Oxford University Press.



\bibitem{sym} I.~F.~Ginzburg, M.~Krawczyk, \emph{Symmetries of Two Higgs Doublet Model and CP violation}, Phys.~Rev.~D 72 (2005) 115013;\\
I.F.~Ginzburg, M.~Krawczyk, \emph{Symmetries in two-Higgs doublet model, CP violation and heavy Higgs effects}, Proceedings of the XVIII International Workshop QFTHEP'2004  (http://theory.sinp.msu.ru/~qfthep04/2004/Proceedings04.html).
\bibitem{ma} N.~G.~Deshpande, E.~Ma, \emph{Pattern of symmetry breaking with two Higgs doublets}, Phys.~Rev.~D 18 (1978) 2574-2576.

\bibitem{sm-like} I.~F.~Ginzburg, M.~Krawczyk, P.~Osland, \emph{Resolving SM-like scenarios via Higgs boson production at a Photon Collider: I. 2HDM versus SM}, IFT 2000-21, arXiv:hep-ph/0101208.



\bibitem{lqt} B.~W.~Lee, C.~Quigg, H.~B.~Thacker, \emph{Weak interactions at vary high energies: The role of the Higgs-boson mass}, Phys.~Rev.~D 16 (1977) 1519;\\
B.~W.~Lee, C.~Quigg, H.~B.~Thacker, \emph{Strength of Weak Interactions at Very High Energies and the Higgs Boson Mass}, Phys.~Rev.~Lett. 38 (1977) 883.




\bibitem{tanb} V.~D.~Barger, J.~L.~.Hewett, R.~J.~N.~Phillips, \emph{New constraints on the charged Higgs sector in two-Higgs-doublet models}, Phys.~Rev.~D 41 (1990)  3421.




\end{thebibliography}
\end{document}